\documentclass[prb, twocolumn,showkeys]{revtex4-1}

\usepackage{amsmath,amssymb,bm}
\usepackage{hyperref}
\usepackage{graphicx}
\usepackage{epstopdf}
\usepackage{latexsym}
\usepackage[usenames, dvipsnames]{color}
\usepackage{natbib}
\usepackage{braket}
\usepackage{float}
\usepackage[normalem]{ulem}
\usepackage{comment}
\usepackage{mathtools}
\usepackage{array}
\usepackage{tabu}
\usepackage{multirow}

\usepackage{color}
\usepackage{lipsum}
      
\usepackage{hyperref}
\hypersetup{
  colorlinks,
  citecolor=magenta,
  linkcolor=blue,
  urlcolor=blue}

\begin{document}

\title{Stability of zero energy Dirac touchings in the honeycomb Hofstadter problem}

\author{Ankur Das}
\author{Ribhu K. Kaul}
\author{Ganpathy Murthy}
\affiliation{Department of Physics \& Astronomy, University of Kentucky, Lexington, KY 40506-0055}

\begin{abstract}
We study the band structure of electrons hopping on a honeycomb
lattice with $1/q$ ($q$ integer) flux quanta through each elementary hexagon. In the
nearest neighbor hopping model the
two bands that eventually form the $n=0$ Landau level have $2q$ zero energy Dirac
touchings. In this work we study
the conditions needed for these Dirac points and their stability to various perturbations. We prove that these touchings and
their locations are guaranteed by a combination of  an anti-unitary
particle-hole symmetry and the lattice symmetries of the
honeycomb structure. We also study the stability of the Dirac
touchings to one-body perturbations that explicitly lower the symmetry.
\end{abstract}
\date{\today}
\maketitle

\section{Introduction}

The band structure of electronic energy levels is a fascinating
consequence of quantum mechanics applied to a
solid.~\cite{ashcroft76:am} The study of the topology of band
structures has received tremendous attention in the last decade
highlighted by the discovery of a variety of topological insulators
and nodal semi-metals.~\cite{KaneMele2005,KaneMele20052,MooreBalents2007,FuKaneMele2007,FuKane2007,roy09:z2,hasan2010:ti} An important theme that has
emerged is the importance of symmetries in protecting the distinction
between insulating states and also the gaplessness of semi-metals.~\cite{DiracPtStability2007,Murakami2007,Vishwanath2016,turner13:sm}

The study of the linear band touching in graphene~\cite{DiracPtStability2007,Hou4} has played a
profound role in the unfolding of these discoveries. It is now well
known that any tight-binding model of graphene (our discussion here
will ignore both spin-orbit coupling and electron-electron
interactions) with time-reversal symmetry and the symmetry of the
honeycomb lattice has two independent Dirac touchings in its Brillouin
zone at the K and K$^\prime$ points. These touchings are stable to a
number of quadratic perturbations. If the translational symmetry of
the Bravais lattice is preserved, the touchings are stable to any
perturbation that preserves rotation by $\pi$ around the honeycomb
center and time reversal. Such perturbations can cause the Dirac
touchings to move in the BZ, but they cannot gap them out.  Breaking
inversion or time-reversal symmetry individually lead to a trivial or
Chern insulator respectively. A periodic perturbation that breaks the
translational symmetry of the original Bravais lattice with a
wavevector that connects the K and K$^\prime$ points can also gap the
Dirac points out (e.g. a Kekule dimerization). The derivation of these
results is reviewed in Appendix~\ref{app:zerofg}.

Our goal in this paper is to generalize these results to the honeycomb
lattice in a magnetic field. This leads to integer quantum Hall
states, historically the first examples of topological states of
matter\cite{TKNN}, which in turn inspired the construction of the
first lattice model of a Chern band.~\cite{Haldane88}

In this work we study the $2q$ Dirac touchings that arise in the
central two bands of the nearest-neighbor tight-binding honeycomb
lattice when a flux of $1/q$ ($q$ integer) times the flux quantum is
introduced into each elementary honeycomb plaquette -- the so-called
Hofstadter problem.~\cite{hofstadter76:bf} There has been quite a bit
of work on the Hofstadter problem on the honeycomb lattice. For the
nearest-neighbor hopping model, previous work has, among other things,
studied the spectrum and the
eigenstates~\cite{Rammal,Kohmoto2006,Agazzi2014}, the Diophantine
equation and Chern number characterizing gapped states\cite{Satoetal},
the crossover from Dirac-like behavior to conventional nonrelativistic
behavior~\cite{Hatsugaietal2006,Bernevigetal2006}, and the approach to
the continuum limit $q\to\infty$.~\cite{Kohmoto2006,Bernevigetal2006}
The existence of $2q$ Dirac band touchings of the central two bands in the
nearest-neighbor hopping model was noticed by several authors, and
explored thoroughly more recently.~\cite{rhim2012:dirac} It was also
pointed out that adding a next-nearest neighbor hopping gaps the Dirac
points out.~\cite{karnaukhov}

Here we extend the discussion in two ways: We first prove explicitly
that certain specific symmetries protect the $2q$ Dirac touchings in a
family of hopping models with arbitrary range hoppings, and $1/q$
flux. Second, we study the stability of these linear touchings to
various one-body perturbations that lower the symmetry.

\section{Model}

Throughout this paper we will be interested in the problem of spinless
fermions hopping on the honeycomb lattice in the presence of a uniform
magnetic field. We will study the problem in the tight binding limit
and assume that each unit cell of the honeycomb lattice encloses
a fraction $1/q$ of the flux quantum.

\subsection{Gauge}

We use the following conventions to define our honeycomb lattice. The
two lattice vectors defining the primitive triangular lattice are,
\begin{eqnarray}
\mathbf a_1 &=& a \hat x\\
\mathbf a_2 &=& a \left ( \frac{\hat x}{2} + \frac{\sqrt{3}\hat
             y}{2}\right )
\end{eqnarray}
With these definitions the vectors describing the sites of the
honeycomb lattice are,
\begin{eqnarray}
\mathbf r_\mu (\bf {n})&=& n_1 \mathbf a_1 +n_2 \mathbf a_2+\mu
\frac{a}{\sqrt{3}}\hat y\\
&=& a (n_1+\frac{n_2}{2})\hat x + \frac{\sqrt{3}}{2}a \hat y (n_2 +\frac{2\mu}{3})
\end{eqnarray}
where ${\bf n}=(n_1,n_2)$ is a pair of integers and $\mu=0,1$ for the
A and B sublattice respectively.

\begin{figure}
\centerline{\includegraphics[angle=0,width=1.0\columnwidth]{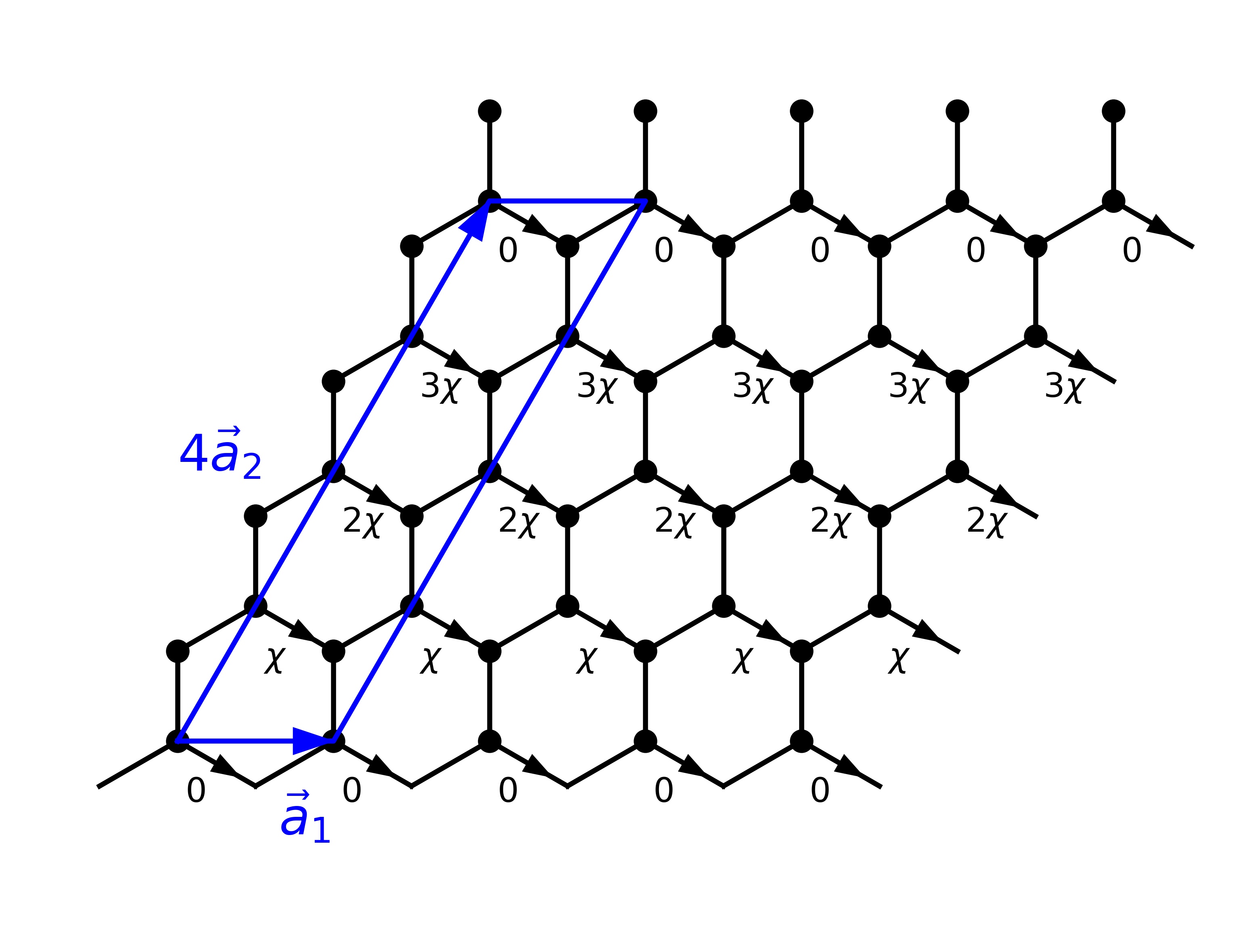}}
\caption{A section of the honeycomb lattice showing the unit cell in
  the optimal gauge (OG) for $q=4$. The magnetic unit cell (MUC),
  defined such that there are two commuting translation operators
  which commute with the Hamiltonian, contains $q$ elementary unit
  cells, and thus $2q$ lattice sites, The lattice vectors associated
  with the MUC are $\mathbf a_1$ and $q \mathbf a_2$. The pattern of
  the phases of the nearest neighbor hoppings in the optimal gauge
  $(0,\chi,2\chi,3\chi)$ are shown ($\chi=\frac{2\pi}{q}$). Additional neighbor hopping can be
  included using the formula, Eq.~(\ref{eq:ppog}) without increasing
  the size of the unit cell.  }
\label{fig:lattog}
\end{figure}

Once we introduce a rational magnetic field
\begin{equation}
  \frac{eB\sqrt{3}a^2}{2\hbar}=\frac{2\pi}{q}\equiv\chi
\end{equation}
the hoppings acquire phases and we have to enlarge our unit cell in
order to obtain two commuting translations, enabling us to apply
Bloch's theorem to compute the band structure.  This enlarged unit
cell is the magnetic unit cell (MUC). From now on we will use $a=1$. It will be useful for us to
start with a continuum gauge, obtain the hopping phases, and then
transform to the final gauge. We begin with the standard Landau gauge,
\begin{equation}
\label{eq:clg}
\mathbf A = -B y {\bf x}
\end{equation}
We introduce the external magnetic field into the hopping model using the
Peierls substitution to calculate the phase of the matrix elements. 
Using this gauge and the standard formula for the Peierls phase
between two lattice points described by ${\bf n},\mu$ and ${\bf n+\Delta n },\nu$:
\begin{eqnarray}
\label{eq:pplg}
\phi^L_{\mu\nu}({\bf n};{\bf \Delta n})&=& \frac{e}{\hbar} \int_{{\bf n},
  \mu}^{{\bf n+\Delta n },\nu} \mathbf A.d\mathbf {l}\\
&=& - \chi \left [ n_2 + \frac{\Delta n_2}{2} +\frac{\mu+\nu}{3}\right
] \left [\Delta n_1 + \frac{\Delta n_2}{2}\right ]\nonumber
\end{eqnarray}
where $\chi \equiv eBa^2 \sqrt{3}/(2\hbar)=2\pi/q$ is the flux per
unit cell of our system (in units of the flux quantum $\frac{h}{e}$),
and ${\bf \Delta n}=(\Delta n_1,\Delta n_2)$. From the expression it
is clear that $\phi^L_{\mu\nu}({\bf n};{\bf \Delta n})$ depends
explicitly on $n_2$ and is $2\pi$ periodic only after $2q$ steps in
the $n_2$ direction. Since $n_1$ does not appear it is periodic in
every step of $n_1$. This means that we need to include $2q$ unit
cells of the triangular lattice in our magnetic unit cell. We can see
this explicitly by constructing the nearest neighbor hopping
Hamiltonian in the Landau gauge,
\begin{eqnarray}
H_{\rm nn} &=& -t\sum_{{\bf n}} d^\dagger_{A,n_1,n_2} [
  d_{B,n_1,n_2}+e^{-i\frac{\chi}{2}(n_2-\frac{1}{6})}d_{B,n_1,n_2-1} \nonumber\\
&+& e^{i\frac{\chi}{2}(n_2-\frac{1}{6})}d_{B,n_1+1,n_2-1}  + {\rm h.c.}],
\end{eqnarray}
which clearly repeats itself with a magnetic unit cell consisting of
$2q$ triangular unit cells.  This is somewhat unsatisfactory since
with a flux of $2\pi/q$ in a triangular unit, a gauge should exist in
which there are only $q$ triangular units in the magnetic unit cells
-- the minimum size of unit cell needed to enclose an integer number
of flux quanta. This can be resolved by working in the so-called
optimal gauge (OG). To achieve this we make the following gauge
transformation from the $d$ fermions (Landau gauge) to a set of $c$
fermions (OG),
\begin{equation}
d_{\mu,n_1,n_2}= e^{-i \frac{\chi}{4}n_2^2+i \frac{\chi}{6}(n_1-n_2)}c_{\mu}(n_1,n_2).
\end{equation}
Using the transformation we can now compute the Peierls phase
between two arbitrary sites on the honeycomb in the OG,
\begin{eqnarray}
\label{eq:ppog}
\phi^{\rm OG}_{\mu\nu}({\bf n};{\bf \Delta n})
&=& - \chi \big [n_2 \Delta n_1 + \frac{2\mu+2\nu+1}{6}\Delta n_1\nonumber \\  &+&
  \frac{\mu+\nu-1}{6}\Delta n_2
+\frac{\Delta n_1 \Delta n_2}{2}
\big ]
\end{eqnarray}
From this formula, in the OG it is clear that the phases repeat
themselves after $q$ steps in the $n_2$ direction, and thus Bloch's
theorem can be applied with only $q$ units of the triangular lattice
in the magnetic unit cell (which contain 2$q$ lattice sites). We shall
choose the magnetic unit cell shown in Fig.~\ref{fig:lattog} in the
rest of the paper and refer to it as the MUC. We can see the
periodicity of the MUC explicitly by working out the nearest neighbor
Hamiltonian in the OG,
\begin{eqnarray}
\label{eq:Hnnog} 
H_{\rm nn} &=& -t\sum_{{\bf n}} c^\dagger_{A}(n_1,n_2) [
  c_{B}(n_1,n_2)+c_{B}(n_1,n_2-1) \nonumber \\
&+& e^{i\chi n_2}c_{B}(n_1+1,n_2-1)]+ {\rm h.c.},
\end{eqnarray}
which clearly repeats itself after $q$ steps in the $n_2$
direction. The advantage of constructing the OG starting from the
Landau gauge is that we now have a definite prescription to compute the
Peierls' phase for an arbitrary hopping matrix element in this gauge,
Eq.~(\ref{eq:ppog}). This allows us to write down hopping models with
an arbitrary range of hopping such that all close paths enclose
precisely the flux corresponding to a uniform external magnetic field,
all the while still
maintaining the MUC containing 2$q$ sites.

\section{Dirac touchings}

Working in the OG we have computed the band structure for
various ranges of tight binding models. This involves the
diagonlization of a $2q \times 2q$ matrix for each ${\bf
  k}$ in the first Brillouin zone. The unit cell we have chosen and
other lattice conventions are shown in Fig.~\ref{fig:lattog}.

\begin{figure}[!t]
  \centering 
  \includegraphics[width=0.95\linewidth]{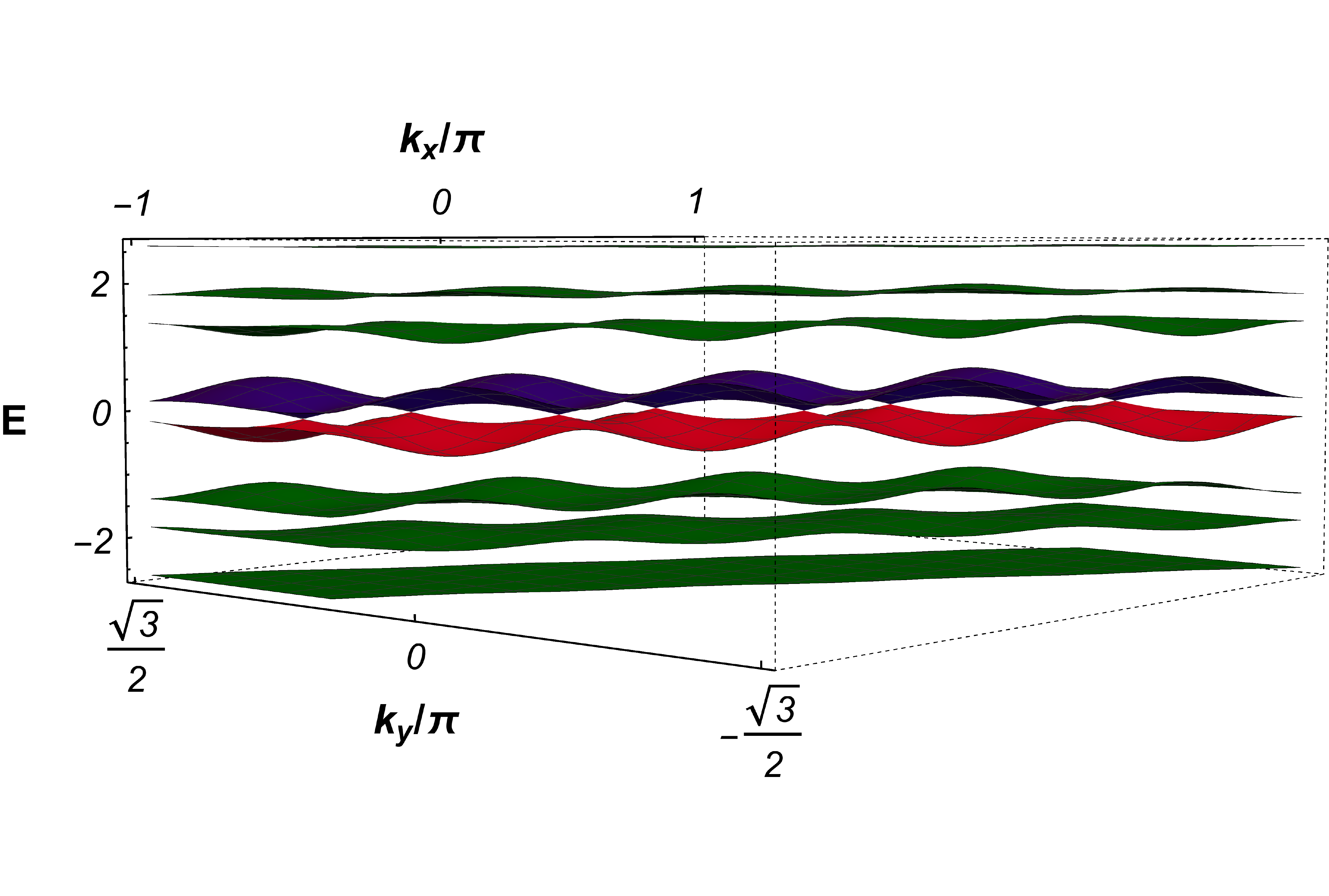}  
  \label{fig:q4bs}
  \includegraphics[width=0.95\linewidth]{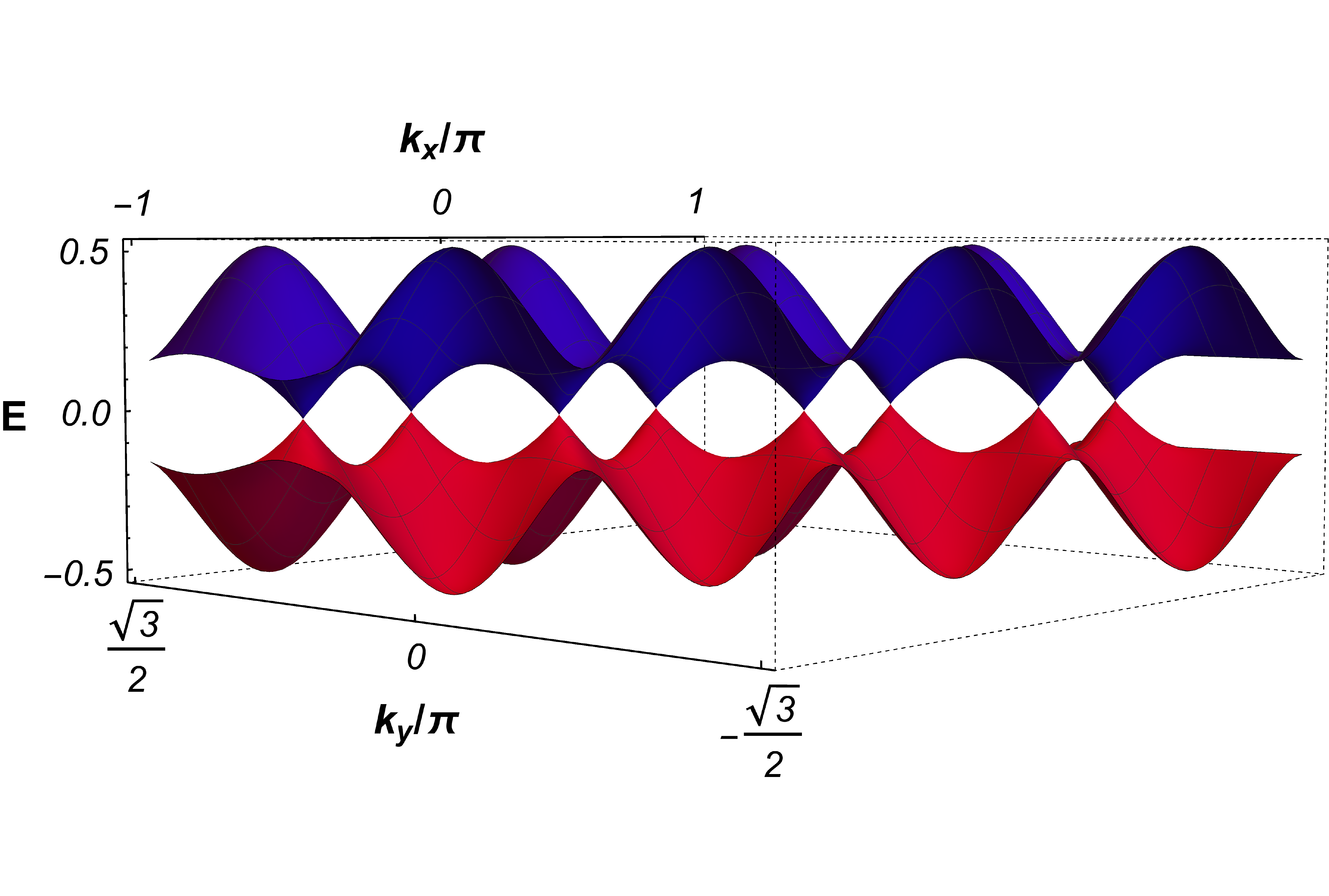}  
  \label{fig:q4bsz}
  \includegraphics[width=.95\linewidth]{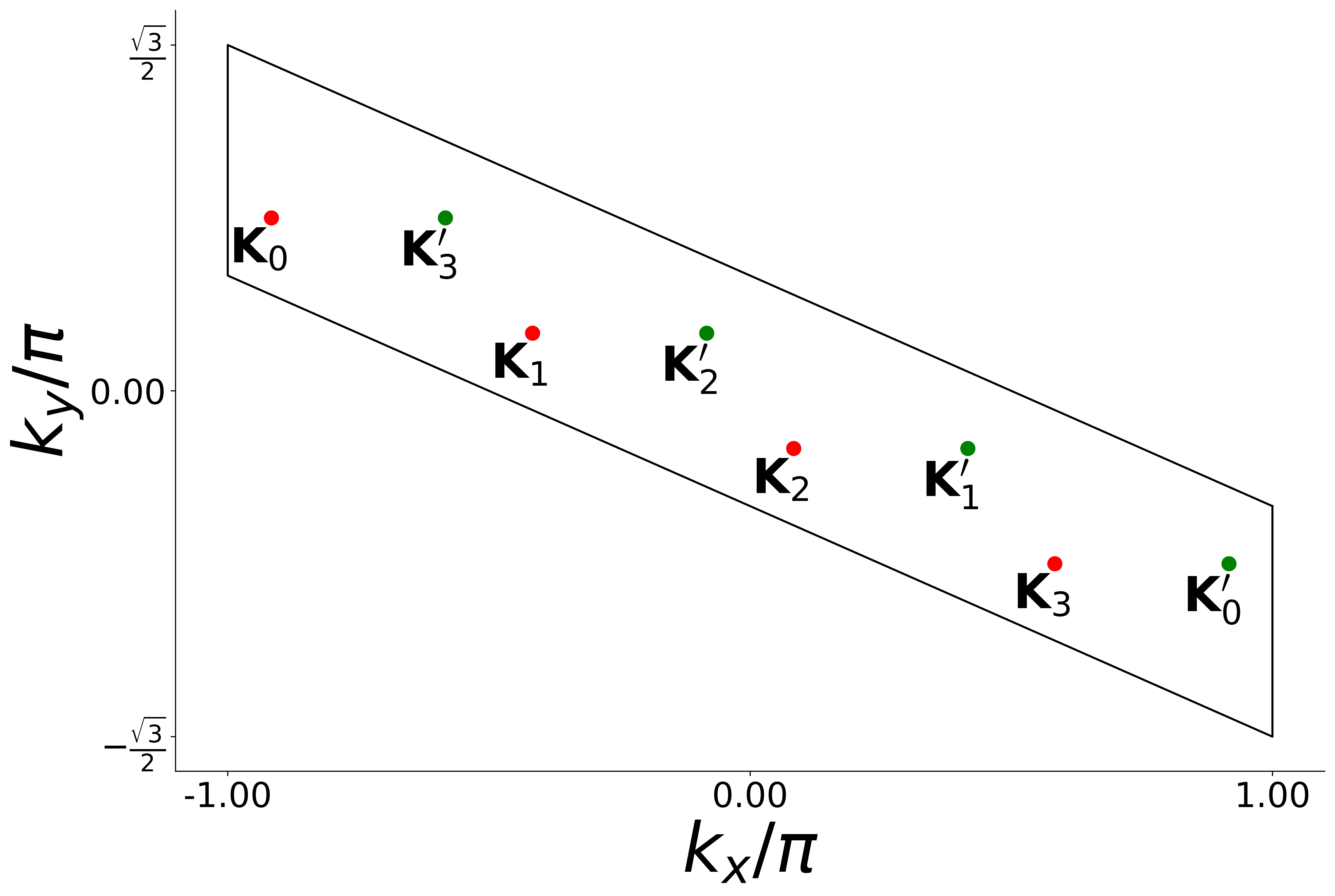}  
  \label{fig:q4bz}
\caption{Band structure  in the first Brillouin zone of the $q=4$ problem with only the nearest
  neighbor hopping. The top panel shows all $2q=8$ bands. The middle
  panel shows the central two bands which touch at linear
  Dirac crossings 2$q$ times -- these bands are the focus of the study
  here, they form the $n=0$ Landau level of graphene in the continuum
  limit. The location of the Dirac points in the BZ corresponding to
  the MUC shown in Fig.~\ref{fig:lattog} are marked
  in the lower panel. }
\label{fig:q4bands}
 \end{figure}

Fig.~\ref{fig:q4bands} shows the electronic structure of the nearest
neighbor model with $q=4$.  Our focus in this paper is on the
finite-$q$ electronic structure of the two central Bloch bands that
eventually form the zero energy $n=0$ continuum Landau levels. In
particular, as has been noticed in previous work, for the
nearest-neighbor model the two bands have $2q$ linear band touchings
that form a honeycomb lattice in reciprocal
space.~\cite{rhim2012:dirac} As $q$ is increased the bandwidth of
these bands decreases exponentially~\cite{Bernevigetal2006} and
eventually as $q\rightarrow \infty$ we recover dispersionless Landau
levels.

Are these Dirac touchings special to the nearest neighbor model or are
they generic to the inclusion of further neighbor hoppings? It is
known~\cite{Karnaukhov2019} that a next-nearest neighbor hopping gaps
out the Dirac points. The formula Eq.~(\ref{eq:ppog}) in the optimal
gauge consistently allows us to include any range of hopping in the
presence of a uniform field. We shall prove below that the Dirac
points and their location are stable as long as the further neighbor
hoppings are bipartite, i.e. they only connect sites on A with sites
on B and maintain spatial the symmetries of the honeycomb lattice. If
any A-A and B-B hoppings are included they gap out the Dirac fermions
even if the honeycomb spatial symmetries are maintained.

As will be crucial for our discussion, the bipartite hopping structure
has an extra symmetry that is broken when hopping between same
sublattices is included. We note that in our problem both
time-reversal symmetry $\mathbb{T}$ and the standard bipartite
particle-hole symmetry $\mathbb{C}$ are broken, since physically they
both reverse the direction of the external magnetic field. However,
the product of the two, the sublattice symmetry $\mathbb{S}$ (which is
an anti-unitary many-body particle-hole transformation) commutes with
the Hamiltonian when only bipartite hoppings are included. The
sublattice symmetry $\mathbb{S}$ is distinct from the ``hidden
symmetry''~\cite{Hou1,Hou2,Hou3,Hou4} which exists on certain lattices and
is also antiunitary, but in addition involves a translation and a
sublattice exchange.

We prove explicitly that with the added constraint of the presence of
$\mathbb{S}$ (in addition to all the lattice symmetries of honeycomb
graphene lattice) all hopping models in the presence of a uniform
magnetic field on the honeycomb lattice have $2q$ Dirac touchings at zero
energy at the same locations in the BZ as the nearest neighbor
model. We have tested this assertion by numerical
diagonalization for a variety of different choices of the range and
magnitude of the hoppings.

\subsection{Proof of Dirac touching at special points}
\label{proof}

We will prove, by contradiction, that for generic Hamiltonians
preserving the lattice and $\mathbb{S}$, there are necessarily
zero-energy states at the 2$q$ special points in the BZ (the same
points for which the nearest neighbor model has Dirac touchings).

Let us briefly
introduce the action of various symmetries on the fermion operators in
the OG (A more detailed discussion is presented in the appendices). Here $(n_1,n_2)$ are the two integers that
label the location of a Bravais lattice sites of the original unit
cell ({\it not} the magnetic unit cell), $\mu,\nu=0,1$ label the $A$
and $B$ sublattices. The two translations in the $\mathbf{a}_1$ and
$\mathbf{a}_2$ directions act as follows:
\begin{subequations}
    \begin{align}
        \mathbb{T}_{\mathbf{a}_1}c_\mu(n_1,n_2)\mathbb{T}^\dagger_{\mathbf{a}_1}&=c_\mu(n_1+1,n_2),\\
        \mathbb{T}_{\mathbf{a}_2}c_\mu(n_1,n_2)\mathbb{T}^\dagger_{\mathbf{a}_2}&=e^{i\chi n_1}c_\mu(n_1,n_2+1),
    \end{align}
\end{subequations}
A rotation by $\frac{2\pi}{3}$ around the $A$ site at $n_1=n_2=0$,
$\mathbb{R}_{\frac{2\pi}{3}}$, acts as follows:
\begin{equation}
    \begin{aligned}
        &\mathbb{R}_{2\pi/3}c_\mu(n_1,n_2)\mathbb{R}_{2\pi/3}^\dagger\\
        &~~~~~~~~=e^{i\chi\left(m_1m_2+\frac{m_2(m_2-1)}{2}\right)}c_\mu(m_1,m_2-\mu),
    \end{aligned}
\end{equation}
where $m_1=-n_1-n_2$ and $m_2=n_1$. A rotation by $\pi$ around the
center of a vertical nearest-neighbor $AB$ bond, $\mathbb{R}_{\pi}$,
acts as follows:
\begin{subequations}
    \begin{align}
        \mathbb{R}_\pi c_\mu(n_1,n_2) \mathbb{R}_\pi^\dagger = e^{-i
      \chi n_1}c_{1- \mu}(-n_1,-n_2)
    \end{align}
\end{subequations}
Finally, the  anti-unitary particle-hole symmetry $\mathbb{S}$ acts as follows:
\begin{subequations}
    \begin{align}
        \mathbb{S}c_A(n_1,n_2)\mathbb{S}^{-1}=&c_A^\dagger(n_1,n_2)\\
        \mathbb{S}c_B(n_1,n_2)\mathbb{S}^{-1}=&-c_B^\dagger(n_1,n_2)\\
        \mathbb{S}i\mathbb{S}^{-1}=&-i
    \end{align}
\end{subequations}
As has been noticed in previous work the nearest-neighbor only hopping
model has Dirac touchings in the central two bands. We reproduce the
locations, labelled by $n=0,\dots,q-1$, from Appendix~\ref{ap:oddq}
here for convenience. The $q$ $K$-type points, for $q$ odd, are
\begin{equation}
  \mathbf{K}_n=\frac{\pi}{q}\bigg(2n-q+\frac{1}{3}\bigg){\hat{x}} -\frac{\pi}{q\sqrt{3}}(2n-q-1){\hat{y}}
\end{equation}
with the  $q$ $\mathbf{K}^\prime$-type points being $\mathbf{K}^\prime_n=-\mathbf{K}_n$.

Here are some properties of the $\mathbf{k}$-space Hamiltonians at these
points that we will need. The details are in the appendices.

{\bf P1}: The
translation operator $\mathbb{T}_{\mathbf{a}_2}$ sends the Hamiltonian at $\mathbf{K}_n$ to the
Hamiltonian at $\mathbf{K}_{n-1}$ (mod $\mathbf{G}_1$). Similarly for the
points $\mathbf{K}^\prime_{n}$. Since the real-space Hamiltonian commutes with
$\mathbb{T}_{\mathbf{a}_2}$, the spectrum must be identical at all the $\mathbf{K}_n$ points.

{\bf P2}: The rotation $\mathbb{R}_\pi$ takes the set of $\mathbf{K}_n$
points to the set of $\mathbf{K}^\prime_n$ points. Since this is a symmetry of
the Hamiltonian, the spectrum at the $\mathbf{K}^\prime_n$ points is identical
to that at the $\mathbf{K}_n$ points. Together with {\bf P1}, this means
that it is sufficient to understand the spectrum at a single
$\mathbf{K}_n$ point.

{\bf P3}:  A rotation by $\frac{2\pi}{3}$ of the
destruction operator at an arbitrary point $\mathbf{k}$ in the BZ leads
to a superposition of destruction operators at the points
\begin{equation}
  \mathbf{k}_{\gamma}=\mathbf{k}_R+\mathbf{G}_2\frac{q+1}{2q}+\gamma\frac{\mathbf{G}_1}{q}
\end{equation}
where $\mathbf{k}_R$ is simply $\mathbf{k}$ geometrically rotated by
$\frac{2\pi}{3}$, and $\gamma=0\dots,q-1$. $\mathbf{G}_1$ and $\mathbf{G}_2$ are the reciprocal lattice vectors of the original lattice. The set of points
$\mathbf{K}_n$ are taken into each other by this transformation, as are
the points $\mathbf{K}^\prime_n$. The operator transformations for the fermion
operators can be found in Appendix C.  Note that
$\mathbb{R}_{\frac{2\pi}{3}}$ preserves the sublattice index.

{\bf P4}: The chiral symmetry $\mathbb{S}$ means that at any point
$\mathbf{k}$ in the BZ the Hamiltonian can be written in block form
\begin{equation}
\label{off-diag-form}
  H(\mathbf{k})=\left(\begin{array}{cc}
  0&M\\
  M^\dagger&0
  \end{array}\right)
\end{equation}

where the q $A$-type sublattice sites have been listed first, and the
q $B$-type sublattices have been listed next. 

Now we are ready for the proof by contradication. Let us assume that
there are no zero-energy states at a particular $\mathbf{K}_n$ point. Let
us further assume that there are no degneracies in the spectrum, so
there are $2q$ nondegenerate states. 

Eq. (\ref{off-diag-form}) implies two facts. Firstly, any eigenstate
of energy $E\neq0$ is necessarily a superposition of $A$ and $B$
sublattices $[\psi_A,\psi_B]^T$, with non-zero amplitudes on
both. Secondly, for every eigenstate with energy $E\neq0$, there is
another eigenstate $[\psi_A,-\psi_B]^T$ with energy $-E$. The
orthogonality of these two eigenstates implies that each $E\neq 0$
eigenstate has equal probabilities on the $A$-type and $B$-type
sublattices.

Let us consider an eigenstate of $H(\mathbf{K}_n)$ at a particular
$n$. For notational convenience we will drop the $\mathbf{K}$ in what
follows, and refer to objects at $\mathbf{K}_n$ simply by the subscript
$n$, e.g., $c_{A\alpha}(\mathbf{K}_n)\equiv  c_{A,\alpha,n}$. By assumption
the eigenstate we consider has $E\neq0$. We can write the destruction operator for this
eigenstate as
\begin{equation}
\label{eneq0state}
  f_n(E)=\sum\limits_{\alpha=0}^{q-1}\bigg(\psi^{(E)}_{A,\alpha,n}c_{A,\alpha,n}+\psi^{(E)}_{B,\alpha,n}c_{B,\alpha,n}\bigg)
\end{equation}

Now we apply the $\mathbb{R}_{\frac{2\pi}{3}}$ to this equation.
Since $[\mathbb{R}_{\frac{2\pi}{3}},H]=0$, the result will be a
superposition of operators corresponding to eigenstates at the same
energy $E$ at all the $\mathbf{K}$-type points. 
\begin{equation}
\mathbb{R}_{\frac{2\pi}{3}}f_n(E)\mathbb{R}^{\dagger}_{\frac{2\pi}{3}}=\sum\limits_{n'=0}^{q-1}|t_{nn'}|e^{i\phi_{nn'}}f_{n'}(E)
\end{equation}

Now focus on the $n=n'$ term on the RHS. From {\bf P3} we know that
$\mathbb{R}_{\frac{2\pi}{3}}$ does not mix the $A$ and $B$
sublattices. Thus, the restriction of $\mathbb{R}_{\frac{2\pi}{3}}$ to
$n=n'$ is a block diagonal $2q\times2q$ matrix.
\begin{equation}
  \langle\mu,\alpha,n|\mathbb{R}_{\frac{2\pi}{3}}|\nu,\beta,n\rangle =\left(\begin{array}{cc}
  R_A(n)&0\\
  0&R_B(n)
  \end{array}\right)
\end{equation}
where each of $R_{A}(n)$ and $R_B(n)$ are $q\times q$ matrices. 

Applying this to Eq. (\ref{eneq0state}) we see that
$\psi_{A,\alpha,n}$ must be an eigenstate of $R_A(n)$, and
$\psi_{B,\alpha,n}$ must be an eigenstate of $R_B(n)$, {\it with the
  same eigenvalue}. Note that if an eigenstate of $H(n)$ had zero
energy, it need not have nonzero amplitudes in both $A$ and $B$
sublattices, and so could evade this conclusion.

By assumption, all the eigenstates have nonzero energy. Thus, {\it all the
eigenvalues of $R_A(n)$ and $R_B(n)$ must be identical}. This leads to
the conclusion that
\begin{equation}
\label{det-real}
  det\bigg( R_A(n)\big(R_{B}(n)\big)^\dagger\bigg)=real
\end{equation}

From the explicit forms of $R_A(n)$ and $R_B(n)$ obtained by
restricting the expressions in Appendix C one easily obtains 
\begin{equation}
q  \bigg(R_A(n)\big(R_{B}(n)\big)^\dagger\bigg)_{\alpha\beta}=e^{i\frac{\chi}{3}+i\chi\beta}\delta_{\beta,\alpha-1}
\end{equation}
From this, one sees that
\begin{equation}
  arg\bigg( det\bigg( R_A(n)\big(R_{B}(n)\big)^\dagger\bigg)\bigg)=\frac{2\pi}{3}
\end{equation}
\label{det-not-real}

This contradicts our conclusion of Eq.~(\ref{det-real}).  Thus, at
least some of the states at $\mathbf{K}_n$ must have zero energy. From
the fact that the chiral symmetry implies that energies must occur in
pairs of $\pm E$, an even number of states must have zero energy at
any $\mathbf{K}_n$.

This shows that there must be band touching at the $\mathbf{K}_n$ and
$\mathbf{K}^\prime_n$ points. Carrying out $\mathbf{k}\cdot\mathbf{p}$ perturbation
theory around a $\mathbf{K}_n$ point, there is no symmetry reason for the
first derivative to vanish, and thus the touchings will generically be linear. 
This completes our proof.

\section{Stability to Small Perturbations}

In the previous section we have shown that with the graphene lattice
symmetry and $\mathbb{S}$, we have $2q$ Dirac nodes at the specific
locations: $\{\mathbf{K}_0,\cdots,\mathbf{K}_{q-1},\mathbf{K}^\prime_0,\cdots,\mathbf{K}^\prime_{q-1}\}$ at zero energy. We now study the stability of these
Dirac touchings to quadratic perturbations.  Before turning to
specific perturbations, we address this question in more general
topological terms.~\cite{Schnyderetal2008,Kitaev:2009mg} We know that
time-reversal symmetry ($\mathbb{T}$) and particle-hole symmetry
($\mathbb{C}$) are both individually absent, but the composite of the
two, the anti-unitary particle hole ($\mathbb{S}$), is present. A
band insulator with the symmetry $\mathbb{S}$ would be in class
AIII. Our model, with all the symmetries intact, has Dirac touchings
and is thus not a band insulator.  A band insulator in which
$\mathbb{S}$ is broken (say by the introduction of same sub-lattice
hopping or a sub-lattice energy difference) would be in class A. It is
now understood that the stability of Dirac touchings can be explained
by the classification of band insulators in one lower
dimension.~\cite{RyuHatsugai2002} The argument relies on considering
the topological classification of the band insulator Bloch
wavefunction on a ring surrounding the Dirac point.~\cite{turner13:sm} In one dimension
band insulators in class AIII have a $\mathbb{Z}$ classification while
in class A they have only trivial band structures. This integer
winding number can be computed for each Dirac node by using a simple
prescription. Referring to Eq.~(\ref{off-diag-form}), one computes
the winding number of the phase of the determinant of $M$ on a contour in
the BZ around the band touching. The expression for the one dimensional winding number is\cite{Schnyderetal2008,Ryu2010,fulga2012},
\begin{equation}
\mathcal{Q}(H)=\frac{1}{2\pi i}\int_0^{2\pi} d\theta \nabla_\theta ln \det M(\theta)
\end{equation}
Computing the winding number for the Dirac touchings we find that all
$\mathbf{K}$ point have -1 and $\mathbf{K}^\prime$ have +1. That all
of the $\mathbf{K}$ points have equal winding numbers follows from
$\mathbb{T}_{\mathbf{a}_2}$ symmetry and that $\mathbf{K}^\prime$ have
the opposite winding follows from the action of the $\mathbb{R}_\pi$ symmetry operation.

From these general topological considerations, we reach the following conclusions for the stability to
perturbations:

Generically if $\mathbb{S}$ is broken
the Dirac touchings get gapped (at least the argument above does not
guarantee stability -- below we study a few examples numerically to
verify this). The resulting insulator will be in class A, with an
integer Chern number.

What perturbation can open a gap if we preserve $\mathbb{S}$? If the perturbation fits in the MUC
(so that the BZ is unchanged) the Dirac
touchings are stable.  We note that locations in the BZ may move if
the perturbations reduce the symmetry operations from those present in
an undistorted honeycomb lattice. Generally, if we preserve
$\mathbb{S}$, small perturbations can open up the gap only if they are
at a wavevector that connects Dirac points with opposite winding
number. Then in the new smaller BZ (corresponding to the enlarged unit
cell), opposite winding number Dirac points lie on top of each
other. Encircling such double touchings will give no winding number,
invalidating the topological argument for their protection.

We now consider specific lattice examples in which we can study how the
Dirac equation gets gapped.

\begin{figure}
\centerline{\includegraphics[angle=0,width=0.95\columnwidth]{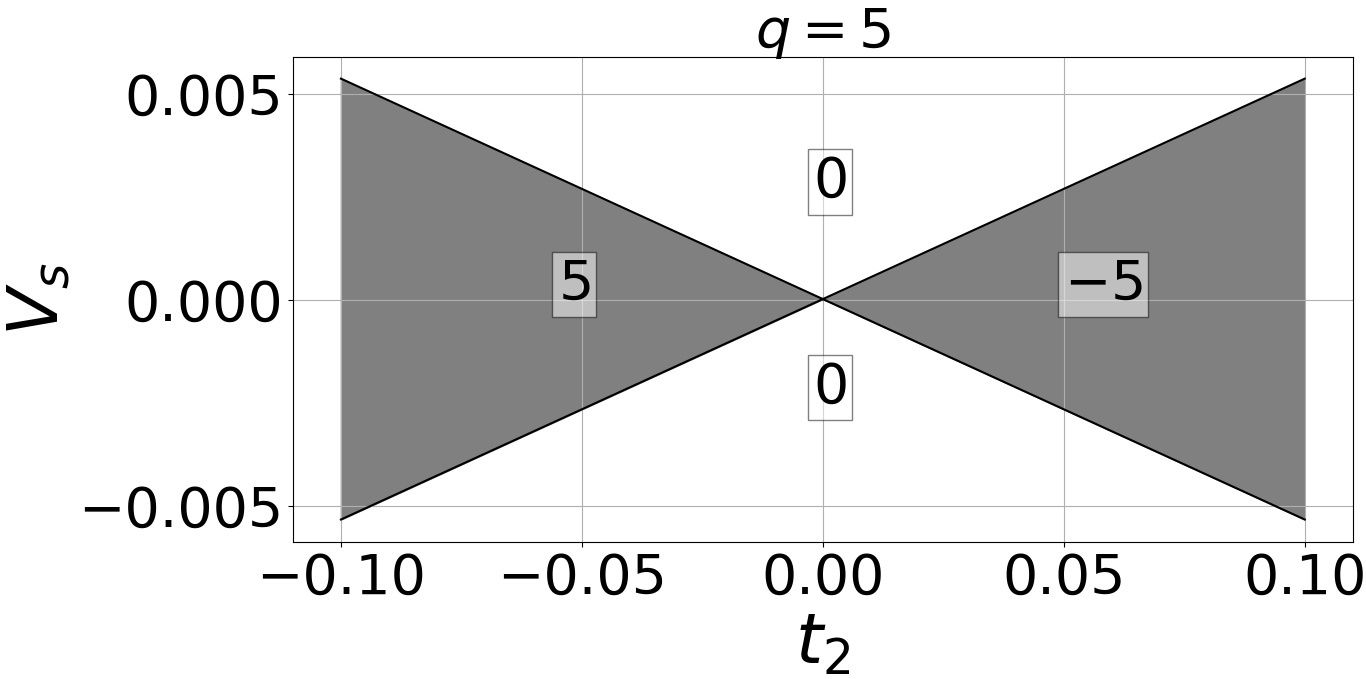}}
\caption{ Phase diagram for $q=5$ showing the Chern numbers of the two
  bands obtained once the Dirac touchings get gapped out. The origin
  corresponds to the nearest neighbor model which has 2$q$ Dirac
  touchings. The Chern numbers of two bands get a uniform contribution
  of (-1,-1) from a Berry curvature distributed throughout the
  BZ. They get an additional contribution from the gapping of the
  Dirac cones, which is sharply localized at these points. $V_s$
  creates a contribution that cancels between the $K$ and $K^\prime$
  Dirac nodes resulting in a net Chern number from only the uniform
  part, i.e.  $(-1,-1)$ and the total Chern number (including all
  occupied bands) at half filling becomes $0$. $t_2$ on the other hand
  creates a net contribution from the gapped Dirac points of $(\pm
  5,\mp 5)$ [generally $(\pm q,\mp q)$] that results in the (4,-6) and
  (-6,4) Chern numbers and the total Chern number at half filling
  becomes $\pm q$ in this case $\pm5$. }
\label{fig:Vst2}
\end{figure}

\subsection{ ${\mathbb S}$-breaking }

We first restrict our discussion to perturbations that
preserve the MUC. To gap the Dirac touchings requires us to break
the $\mathbb{S}$ symmetry. The simplest way to do this is to perturb with a staggered diagonal
energy term in the Hamiltonian that has the same magnitude but
opposite signs on the each of the sub-lattices.  $H =
V_s\sum_{n_1,n_2}\left ( d^\dagger_{A, n_1,n_2}d_{A,
  n_1,n_2}-d^\dagger_{B, n_1,n_2}d_{B, n_1,n_2}\right )$. Although
this fits in the MUC it breaks some of the lattice symmetry,
e.g. $\mathbb{R}_\pi$. A second way to break $\mathbb{S}$ symmetry 
is to include a same sublattice hopping with fixed range for all
sites, e.g. a second neighbor hopping, $t_2$. We include it here with
the correct phase from Eq.~(\ref{eq:ppog}) corresponding to having a
background uniform $B$-field. This perturbation has the feature of
preserving every symmetry in the nearest neighbor hopping model except
for $\mathbb{S}$.  From the arguments made earlier, both perturbations are expected to
open up a gap in the Dirac equation, leaving behind a 2-D insulating
band structure in class A, characterized by an integer Chern
number. 

Since the two middle bands become the $n=0$ Landau level of
graphene, we expect them to have a combined Chern number of -2. How
this -2 is distributed between the two bands obtained after the gap
opening perturbation is added depends on the details. The sign of the
mass that gaps out a particular Dirac point also determines the
transfer of Chern density between the two bands.  Perturbations that preserve translations can only realize the total Chern numbers
$C=0,-q,q$ because the $\mathbb{T}_{\mathbf{a}_2}$ symmetry forces
the form of the Bloch Hamiltonian at all the $\mathbf{K}_n$ points to be the same, and
also all the $\mathbf{K}^\prime_n$ points to be the same. The perturbation
$V_s$ results in a trivial insulator, while the other two values of
$C$ are realized by the $t_2$ perturbation. We have checked all the above assertions by computing the
integer invariant numerically, i.e. by integrating the Berry curvature over the
Brillouin zone. This is shown and discussed in Fig.~\ref{fig:Vst2}.

Above we have studied two examples of perturbations that create
different Chern numbers, 0 and $\pm q$. The Chern numbers of the bands
produced can be any of the intermediate values $1,2\cdots q-1$ as well. This requires
a perturbation that breaks  $\mathbb{T}_{\mathbf{a}_2}$ though it may
preserve the MUC.

\subsection{ ${\mathbb S}$-preserving}

To gap out the Dirac nodes with $\mathbb{S}$ preserved, the
perturbation must break translational invariance with a momentum that
connects Dirac touchings of opposite winding number. The simplest way to
achieve this is to include as a perturbation a periodic modulation of
the magnitude of the first neighbor hopping, with period corresponding
to the {\bf Q}-vector connecting the Dirac nodes. Since there are $q$
nodes with positive winding and $q$ with negative winding, there
appear to be $q^2$ different possibilities. However, only $q$
different {\bf Q}-vectors fit within the magnetic Brillouin zone
leading to $q$ different reduced lattice periodicities. Once the nodes
are gapped out and in the presence of ${\mathbb S}$ we end up with a
band insulator in class AIII. Since these are all trivial insulators
they are expected to be smoothly connected to each other without a gap
closing.

\section{Conclusions}

In conclusion we have studied the stability of the Dirac touchings in
the $n=0$ Landau level in the Hofstadter limit when the external
magnetic field is very strong, with $\frac{1}{q}$ quanta of flux
going through each hexagon.

We started by deriving a
formula in the optimal gauge for an arbitary range hopping, so that
our problem always fits with $q$ unit cells of the honeycomb.
Next, we have shown that the Dirac touchings
require the sublattice symmetry ${\mathbb S}$ for their
protection. Indeed we have proven that every tight binding model with
${\mathbb S}$, the correct flux and with the entire lattice symmetry
of graphene intact will have $2q$ Dirac touchings at the same location
as the nearest neighbor model.

Next, we considered perturbations to the $2q$ Dirac touchings. We
showed from general topological stability arguments as well as
specific hopping models that perturbatively breaking $\mathbb{S}$ or
including a periodic potential that connects Dirac nodes with opposite
winding number can gap the Dirac nodes out. All other perturbations
preserve the Dirac nodes (though their location in the BZ may move). Of
course if these perturbations are made large enough, some finite value
of the perturbation may cause the gapping out of the linear touchings.

Introducing electron-electron interactions is known to produce a rich
set of symmetry-broken phases in graphene for weak
fields.~\cite{Herbut2006,AliceaFisher,Kharitonov,FeshamiFertig} There
have been a few investigations into interaction effects in the
Hofstadter
regime,~\cite{JungMacDonald,Mishraetal2014,Buividovich2016tgo,Mishraetal2016,Mishraetal2017}
but a full picture remains to be developed.

GM thanks  David Mross for a vauable discussion. We acknowledge partial financial support from NSF DMR-1611161 (AD and
RKK) and from NSF DMR-1306897 (AD and GM). We thank the hospitality of
the Aspen Center for Physics (NSF grant no. 1607611) where this work
was finalized for publication. GM also thanks the Gordon and Betty
Moore Foundation for sabbatical support at MIT, and the Lady Davis
Foundation for sabbatical support at the Technion.

\appendix

\section{Graphene without $B$-field}

\label{app:zerofg}

In discussing the Dirac touchings in graphene with
$\mathbb{T}$,~\cite{DiracPtStability2007} it is useful to think about
the problem in two steps as we have done in our manuscript for the
case when $\mathbb{T}$ is broken by an external magnetic field.

First, it is possible using symmetries to prove that for any hopping model that preserves the symmetry of the honeycomb
lattice and time reversal, there are two independent Dirac nodes at K
and K$^\prime$. The symmetry argument does not rule out the existence
of other additional Dirac nodes in the BZ that may coexist with the
two Dirac nodes mandated by symmetry. {\em Proof:} Write the graphene
Hamiltonian as $h= d_x({\bf k})\sigma_x+d_y({\bf k})\sigma_y+d_z({\bf
  k})\sigma_z$. A combination of ${\mathbb R_\pi}$ and $\mathbb{T}$
establish that $d_z=0$. Finally requiring $R_{2\pi/3}$ in addition
forces $(d_x-id_y)_{({\bf K+q)}}\approx v_F (q_x-iq_y)$ and
$(d_x-id_y)_{({\bf K^\prime+q)}}\approx v_F (-q_x-iq_y)$ at leading
order. Using $\mathbf \tau$ as the valley Pauli matrix we obtain the low
energy Hamiltonian as $h\approx \tau_z \sigma_x q_x+\sigma_y q_y$.

Now we turn to the perturbative stability of the Dirac touchings at K
and K$^\prime$, when the symmetry is lowered by either breaking
${\mathbb T}$ or some of the honeycomb lattice symmetries. To gap out
the Dirac fermion we need perturbations that generate mass terms that anticommute with both
$\tau_z \sigma_x$ and $\sigma_y$.   If the translational invariance of the triangular Bravais lattice,
${\mathbb R}_\pi$ and $\mathbb{T}$ are present the touchings are
stable to all perturbations, the only changes to the unperturbed $h$
is a movement of the Dirac touchings. Four mass terms can be added:
$\tau_z \sigma_z$ and  $\sigma_z$  break $\mathbb{T}$ and
$\mathbb{R}_\pi$ leading to the Chern insulator and the trivial band
insulator respectively.~\cite{Haldane88} $\tau_x\sigma_x$ and $\tau_y\sigma_x$ are the
last two, they break translational symmetry of the graphene lattice,
which could arise e.g. from Kekule dimerization.

Our goal in  this paper is to carry out this same two step program for
the problem of the honeycomb lattice with $1/q$ flux. The significant
increase in complication arises from the fact that the matrices that describe this problem are now
$2q$ dimensional.

\section{Symmetry Operations in OG}

In this section we study the various symmetries that are present in
Eq.~(\ref{eq:Hnnog}). We will study these operations by asking how the
symmetry operations act on the lattice creation and destruction
operators in the {\em OG}. While the explicit form of the
transformations are gauge dependent, these explicit forms exist in
every gauge.

Lattice symmetries include translations in the $\mathbf a_1$ and $\mathbf
a_2$ directions, rotations by $2\pi/3$ about a site
and rotation by $\pi$ about the center of a vertical bond. This set of four operations
generates all the spatial symmetry operations present in the
Hofstatder problem. We note here that  mirror symmetries (and generally
all improper
rotations) which are present for the honeycomb lattice structure, are broken by the
presence of the Peierls phases, since they reverse the direction of
the magnetic fluxes.
As noted above, because of the presence of the Peierls phases the
lattice operations must be augmented by a gauge transformation from
the naive operations one writes down in the absence of a magnetic field. In the OG they are,
\begin{subequations}
    \begin{align}
        \mathbb{T}_{\mathbf{a}_1}c_\mu(n_1,n_2)\mathbb{T}^\dagger_{\mathbf{a}_1}&=c_\mu(n_1+1,n_2),\\
        \mathbb{T}_{\mathbf{a}_2}c_\mu(n_1,n_2)\mathbb{T}^\dagger_{\mathbf{a}_2}&=e^{i\chi n_1}c_\mu(n_1,n_2+1),
    \end{align}
\end{subequations}
\begin{equation}
    \begin{aligned}
        &\mathbb{R}_{2\pi/3}c_\mu(n_1,n_2)\mathbb{R}_{2\pi/3}^\dagger\\
        &~~~~~~~~=e^{i\chi\left(m_1m_2+\frac{m_2(m_2-1)}{2}\right)}c_\mu(m_1,m_2-\mu),
    \end{aligned}
\end{equation}
where $m_1=-n_1-n_2$ and $m_2=n_1$, and
\begin{subequations}
    \begin{align}
        \mathbb{R}_\pi c_\mu(n_1,n_2) \mathbb{R}_\pi^\dagger = e^{-i
      \chi n_1}c_{1- \mu}(-n_1,-n_2)
    \end{align}
\end{subequations}
It is easy to verify that each of the above four operations commutes
with the Hamiltonian in the OG, Eq.~(\ref{eq:Hnnog}).

Finally a very important symmetry for our purposes present in Eq.~(\ref{eq:Hnnog}) is
an anti-unitary version of the particle-hole symmetry,
\begin{subequations}
    \begin{align}
        \mathbb{S}c_A(n_1,n_2)\mathbb{S}^{-1}=&c_A^\dagger(n_1,n_2)\\
        \mathbb{S}c_B(n_1,n_2)\mathbb{S}^{-1}=&-c_B^\dagger(n_1,n_2)\\
        \mathbb{S}i\mathbb{S}^{-1}=&-i
    \end{align}
\end{subequations}
which is easily seen to commute with $H_{\rm nn}$. We note that the
conventional time reversal operation $\mathbb{T}$ and the conventional
unitary particle-hole $\mathbb{C}$ each reverse the direction of the
magnetic field and are hence are absent as symmetries in the present
problem. We can then understand that since $\mathbb{S}=\mathbb{TC}$
reverses the magnetic field direction twice it appears as a symmetry
of our problem.

\section{Symmetries in $\mathbf{k}$ Space}

We choose the following periodicity conditions on our fermion operators in the Brillouin zone. 
\begin{subequations}
	\begin{align}
		c_{A\beta}(\mathbf{k}+\tilde{\mathbf{G}}_1)&=c_{A\beta}(\mathbf{k}+\mathbf{G}_1)=c_{A\beta}(\mathbf{k})\\
		c_{B\beta}(\mathbf{k}+\tilde{\mathbf{G}}_1)&=c_{B\beta}(\mathbf{k}+\mathbf{G}_1)=e^{i\frac{2\pi}{3}}c_{B\beta}(\mathbf{k})\\
		c_{A\beta}(\mathbf{k}+\tilde{\mathbf{G}}_2)&=c_{A\beta}\left(\mathbf{k}+\frac{\mathbf{G}_2}{q}\right)=e^{-i\beta\chi}c_{A\beta}(\mathbf{k})\\
		c_{B\beta}(\mathbf{k}+\tilde{\mathbf{G}}_2)&=c_{B\beta}\left(\mathbf{k}+\frac{\mathbf{G}_2}{q}\right)=e^{-i\beta\chi-\frac{2\chi}{3}}c_{B\beta}(\mathbf{k})
	\end{align}
\end{subequations}

Real Space translations,
\begin{subequations}
	\begin{align}
		\mathbb{T}_{\mathbf{a}_1}c_{\mu\alpha}(\mathbf{k})\mathbb{T}^\dagger_{\mathbf{a}_1}&=e^{i\mathbf{k}\cdot \mathbf{a}_1}c_{\mu\alpha}(\mathbf{k})\\
		\mathbb{T}_{\mathbf{a}_2}c_{\mu\alpha}(\mathbf{k})\mathbb{T}^\dagger_{\mathbf{a}_2}&=e^{i\mathbf{k}\cdot \mathbf{a}_1+i\frac{\mu\chi}{3}}c_{\mu[\alpha+1]}\left(\mathbf{k}-\frac{\mathbf{G}_1}{q}\right)
	\end{align}
\end{subequations}
where, $[\alpha+1]=(\alpha+1) mod~q$ and $\alpha\in [0,q-1]$.

Rotation by $\pi$ about a bond center,
\begin{subequations}
	\begin{align}
		\mathbb{R}_\pi c_{A\alpha}(\mathbf{k})\mathbb{R}_\pi^\dagger&=e^{-i(\mathbf{k}+\frac{\mathbf{G}_1}{q})\cdot \mathbf{d}}c_{B\alpha'}\left(-\mathbf{k}-\frac{\mathbf{G}_1}{q}\right)\\
		\mathbb{R}_\pi c_{B\alpha}(\mathbf{k})\mathbb{R}_\pi^\dagger&=e^{-i\mathbf{k}\cdot \mathbf{d}}c_{A\alpha'}\left(-\mathbf{k}-\frac{\mathbf{G}_1}{q}\right)\\
		\text{Where,~}\alpha'&=(1-\delta_{\alpha,0})(q-\alpha)\text{~and~} d=\frac{\hat{y}}{\sqrt{3}} \nonumber
	\end{align}
\end{subequations}

$\frac{2\pi}{3}$ rotation about an $A$ lattice point mixes multiple $\mathbf{k}$ points,
\begin{equation}
	\mathbf{k}^\prime_\gamma=\mathbf{k}_R+\frac{\mathbf{G}_2}{2q}(q+1)+\gamma\frac{\mathbf{G}_1}{q}
\end{equation}
where $\mathbf{k}_R$ is $\mathbf{k}$ rotated by $\frac{2\pi}{3}$.

For $q$ an odd number, we have:
\begin{subequations}
	\begin{equation}
	\begin{aligned}
		\mathbb{R}_\frac{2\pi}{3}c_{A\beta}(\mathbf{k})\mathbb{R}_\frac{2\pi}{3}^\dagger=\frac{1}{q}\sum_{\gamma,\beta'=0}^{q-1}e^{-i\chi(\gamma+\beta')(\beta+\beta')+i\frac{\chi}{2}\beta'(\beta'-1)}\\
		\times c_{A\beta'}\left(\mathbf{k}_R+\frac{\gamma}{q}\mathbf{G}_1\right)
	\end{aligned}
	\end{equation}
	\begin{equation}
	\begin{aligned}
		\mathbb{R}_\frac{2\pi}{3}c_{B\beta}(\mathbf{k})\mathbb{R}_\frac{2\pi}{3}^\dagger=\frac{1}{q}\sum_{\gamma,\beta'=0}^{q-1}e^{-i\chi(\gamma+\beta'+1)(\beta+\beta'+1)+i\frac{\chi}{2}\beta'(\beta'+1)}\\
		\times e^{-i\chi\frac{\gamma}{3}}c_{B\beta'}\left(\mathbf{k}_R+\frac{\gamma}{q}\mathbf{G}_1\right)
	\end{aligned}
	\end{equation}
\end{subequations}

For $q$ an even number, we have:
\begin{subequations}
	\begin{equation}
	\begin{aligned}
		\mathbb{R}_\frac{2\pi}{3}c_{A\beta}(\mathbf{k})\mathbb{R}_\frac{2\pi}{3}^\dagger=\frac{1}{q}\sum_{\gamma,\beta'=0}^{q-1}e^{-i\chi(\gamma+\beta')(\beta+\beta')+i\frac{\chi}{2}{\beta'}^2}\\
		\times c_{A\beta'}\left(\mathbf{k}_R+\frac{\mathbf{G}_2}{2 q}+\frac{\gamma}{q}\mathbf{G}_1\right)
	\end{aligned}
	\end{equation}
	\begin{equation}
	\begin{aligned}
		\mathbb{R}_\frac{2\pi}{3}c_{B\beta}(\mathbf{k})\mathbb{R}_\frac{2\pi}{3}^\dagger=\frac{1}{q}\sum_{\gamma,\beta'=0}^{q-1}e^{-i\chi(\gamma+\beta'+1)(\beta+\beta'+1)+i\chi\beta'(1+\frac{\beta'}{2})}\\
		\times e^{-i\chi\frac{\gamma-1}{3}}c_{B\beta'}\left(\mathbf{k}_R+\frac{\mathbf{G}_2}{2q}+\frac{\gamma}{q}\mathbf{G}_1\right)
	\end{aligned}
	\end{equation}
\end{subequations}

The expressions for the Dirac points, and the restriction of the rotation operators particular Dirac points, naturally fall into two classes, those for $q$ odd, and those for $q$ even. 

\section{Dirac points and $\mathbb{R}_{\frac{2\pi}{3}}$  for q odd}
\label{ap:oddq}
The Dirac points are,
\begin{equation}
\mathbf{K}_n=\frac{\pi}{q}\left( 2n-q+\frac{1}{3} \right)\hat{x}-\frac{\pi}{q\sqrt{3}}(2n-q-1)\hat{y}
\end{equation}
where, $n\in [0,q-1]$ and $\mathbf{K}^\prime_n=-\mathbf{K}_n$.

\begin{equation}
	\begin{aligned}
		\mathbb{R}_\frac{2\pi}{3}c_{A\beta}(\mathbf{K}_n)\mathbb{R}_\frac{2\pi}{3}^\dagger=\frac{1}{q}\sum_{\gamma,\beta'}&e^{-i\chi(\gamma+\beta')(\beta+\beta')+i\frac{\chi}{2}\beta'(\beta'-1)}\\
		&\times e^{-il_2\chi\beta'}c_{A\beta'}(\mathbf{K}_{n'})
	\end{aligned}
\end{equation}
\begin{equation}
	\begin{aligned}
		\mathbb{R}_\frac{2\pi}{3}c_{B\beta}(\mathbf{K}_n)\mathbb{R}_\frac{2\pi}{3}^\dagger=&\frac{1}{q}\sum_{\gamma,\beta'}e^{-i\chi(\gamma+\beta'+1)(\beta+\beta'+1)+i\frac{\chi}{2}\beta'(\beta'+1)}\\
		&\times e^{-\frac{i\chi\gamma}{3}+\frac{2\pi i l_1}{3}-il_2\chi\beta'-\frac{2il_2\chi}{3}}c_{B\beta'}(\mathbf{K}_{n'})
	\end{aligned}
\end{equation}

where,
\begin{subequations}
	\begin{align}
		l_2(n)&=n-\frac{q+1}{2}\\
		n'(n,\gamma)&=\left[ \frac{q-1}{2}+\gamma \right]\in [0,q-1]\\
		l_1(n)&=\begin{cases}
		0,& \text{if,~}\gamma\leq \frac{q-1}{2}\\
		1,& \text{otherwise}
		\end{cases}
	\end{align}
\end{subequations}

\section{Dirac points and  $\mathbb{R}_{\frac{2\pi}{3}}$ for q even}
\label{ap:evenq}
The location of the Dirac points are
\begin{equation}
\mathbf{K}_n=\frac{\pi}{q}\left( 2n-q+\frac{1}{3} \right)\hat{x}-\frac{\pi}{q\sqrt{3}}(2n-q+1)\hat{y}
\end{equation}
where, $n\in [0,q-1]$ and $\mathbf{K}^\prime_n=-\mathbf{K}_n$. Effect of all other operation remain the same as in the odd $q$ case except for the $2\pi/3$ rotations. 

\begin{subequations}
\begin{align}
	\mathbb{R}_{\frac{2\pi}{3}} c^{(\mathbf{K}_n)}_{A\beta}\mathbb{R}^\dagger_{\frac{2\pi}{3}}=\frac{1}{q}\sum_{\gamma,\beta'=0}^{q-1}e^{-i \chi (\gamma+\beta')(\beta+\beta')+i {\beta'}^2\chi/2} \nonumber \\
	\times e^{-i l_2\chi\beta'} c_{A\beta'}^{(\mathbf{K}_{n'})}\\
	\mathbb{R}_{\frac{2\pi}{3}} c^{(\mathbf{K}_n)}_{B\beta}\mathbb{R}^\dagger_{\frac{2\pi}{3}}=\frac{1}{q}\sum_{\gamma,\beta'=0}^{q-1}e^{-i\chi\left[ (\gamma+\beta'+1)(\beta+\beta'+1)-\beta'(1+\frac{\beta'}{2}) \right]} \nonumber \\
	\times e^{-i\chi\frac{(\gamma-1)}{3}-il_2\chi\beta'-\frac{2i l_2\chi}{3}+\frac{2\pi i l_1}{3}}	c_{B\beta'}^{(\mathbf{K}_{n'})}
\end{align}
\end{subequations}

with,
\begin{subequations}
	\begin{align}
		l_2(n)&=n-\frac{q}{2}+1\\
		n'(n,\gamma)&=\left[ \frac{q}{2}+\gamma \right]\in [0,q-1]\\
		l_1(n)&=\begin{cases}
		0,& \text{if,~}\gamma\leq \frac{q}{2}-1\\
		1,& \text{otherwise}
		\end{cases}
	\end{align}
\end{subequations}

\section{$R_A(n)$ and $R_B(n)$}

The matrix that rotates the wavefunction into itself (multiplying with $\sqrt{q}$ makes it unitary), for $q$ odd and $n=\frac{q-1}{2}$:
\begin{subequations}
\begin{align}
R_A(\beta,\beta')=&\frac{1}{q}e^{-i\chi\beta\beta'-i\frac{\chi}{2}\beta'(\beta'-1)}\\
R_B(\beta,\beta')=&\frac{1}{q}e^{-i\chi \beta(\beta'+1)-i\frac{\chi}{2}\beta'(\beta'+1)-i\frac{\chi}{3}}
\end{align}
\end{subequations}

Similarly, for $q$ even and $n=\frac{q}{2}$:
\begin{subequations}
\begin{align}
R_A(\beta,\beta')=&\frac{1}{q}e^{-i\chi\beta\beta'-i\frac{\chi}{2}\beta'^2-i\chi\beta'}\\
R_B(\beta,\beta')=&\frac{1}{q}e^{-i\chi\left[\beta(\beta'+1)+\frac{\beta'^2}{2}+2\beta' \right]-i\frac{4\chi}{3}}
\end{align}
\end{subequations}

\section{Rhim-Park Wavefunction}
\label{app:rp}
We can solve the wave function for the nearest-neighbor Hamiltonian using the methods used by Rhim and Park \cite{rhim2012:dirac}. Using Bloch's theorem we can say that the wave function must have the form
\begin{subequations}  
	\begin{align}
         \psi_A(n_1,ql_2+\alpha)=e^{i\mathbf{k}\cdot\mathbf{r}_{A\alpha}(n_1,l_2)}\psi_{A\alpha}(\mathbf{k})\\
         \psi_B(n_1,ql_2+\alpha)=e^{i\mathbf{k}\cdot\mathbf{r}_{B\alpha}(n_1,l_2)}\psi_{B\alpha}(\mathbf{k})
    \end{align}
 \end{subequations}
For the zero energy eigenstate from the Hamiltonian we can write,
 \begin{subequations}
 	\begin{align}
 	\begin{aligned}
 		\psi_B(n_1,n_2)&+\psi_B(n_1,n_2-1)\\
 		&+e^{i\chi n_2}\psi_B(n_1+1,n_2-1)=0
 	\end{aligned}\\
 	\begin{aligned}
 		\psi_A(n_1,n_2)&+\psi_A(n_1,n_2+1)\\
 		&+e^{-i\chi (n_2+1)}\psi_A(n_1-1,n_2+1)=0
 	\end{aligned}
 	\end{align}
 \end{subequations}
Thus using recursion relation we can write,
 \begin{subequations}
 	\begin{align}
 		\psi_{A\beta}(\mathbf{k})&=\left\{\prod_{\alpha=0}^\beta\frac{-e^{-i k_2}}{1+e^{-i(k_1+\alpha\chi)}}\right\}\psi_{A0}(\mathbf{k})\\
 		\psi_{B\beta}(\mathbf{k})&=\left\{\prod_{\alpha=0}^\beta-e^{-i k_2}\left(1+e^{i(k_1+\alpha\chi)}\right)\right\}\psi_{B0}(\mathbf{k})\\\nonumber
 	\end{align}
 \end{subequations}
where $k_1=\mathbf{k}\cdot\mathbf{a}_1$ and $k_2=\mathbf{k}\cdot\mathbf{a}_2$. Now from the self-consistency for Bloch functions gives the condition on the $\mathbf{k}$ momentum values at which the zero-energy states exist. The conditions are,

\begin{subequations}
 	\begin{align}
 		&\left\{\prod_{\alpha=0}^{q-1}\frac{-e^{-i k_2}}{1+e^{-i(k_1+\alpha\chi)}}\right\}=1\\
 		&\left\{\prod_{\alpha=0}^{q-1}-e^{-i k_2}\left(1+e^{i(k_1+\alpha\chi)}\right)\right\}=1
 	\end{align}
 \end{subequations}

The solutions forms a honeycomb lattice in the momentum space. They consist of two sets,

\begin{subequations}
 	\begin{align}
 		k_x^I&=-\pi+\frac{2\pi j_1}{q}+\frac{\pi}{3q}\\
 		k_y^I&=\pi\sqrt{3}+\frac{2\pi}{q\sqrt{3}}(2j_2-j_1)-\frac{\pi}{q\sqrt{3}}
 	\end{align}
 \end{subequations}
and
\begin{subequations}
 	\begin{align}
 		k_x^{II}&=-\pi+\frac{2\pi j_1}{q}-\frac{\pi}{3q}\\
 		k_y^{II}&=\pi\sqrt{3}+\frac{2\pi}{q\sqrt{3}}(2j_2-j_1)+\frac{\pi}{q\sqrt{3}}
 	\end{align}
 \end{subequations}
Where $j_1,j_2$ can be any integer.

\section{Symmetry action in the low energy space}

Using the Rhim-Park wave function we can derive the action of the
symmetry operations in the low energy space.

Let us start with translations.

\begin{subequations}
\begin{equation}
\mathbb{T}_{\mathbf{a}_1} d_A(\mathbf{K}_n) \mathbb{T}_{\mathbf{a}_1}^\dagger=e^{i \mathbf{K}_n\cdot\mathbf{a}_1} d_A(\mathbf{K}_n)
\end{equation}
\begin{equation}
\mathbb{T}_{\mathbf{a}_1} d_B(\mathbf{K}_n) \mathbb{T}_{\mathbf{a}_1}^\dagger=e^{i \mathbf{K}_n\cdot\mathbf{a}_1} d_B(\mathbf{K}_n)
\end{equation}
\end{subequations}

\begin{equation}
\mathbb{T}_{\mathbf{a}_2} d_\mu(\mathbf{K}_n) \mathbb{T}_{\mathbf{a}_2}^\dagger=e^{i\left(\mathbf{K}_n\cdot\mathbf{a}_2+\frac{\mu\chi}{3}+\frac{\mu \delta_{n,0} 2\pi}{3}\right)} e^{\phi^{\mathbb{T}_{\mathbf{a}_2}}(\mathbf{K}_n)} d_A(\mathbf{K}_n)
\end{equation}

Where,
\begin{equation}
\phi^{\mathbb{T}_{\mathbf{a}_2}}(\mathbf{K}_n)=\begin{cases}
												\frac{2n + q - 1}{2 q}\pi \text{~~~for $q$ odd}\\
												\\
												\frac{2n + q + 1}{2 q}\pi \text{~~~for $q$ even}
												\end{cases}
\end{equation}

Now, consider the  $\pi$ rotation about center of a vertical bond $\mathbb{R}_\pi$.
\begin{subequations}
\begin{equation}
\mathbb{R}_{\pi} d_A(\mathbf{K}_n) \mathbb{R}_{\pi}^\dagger=e^{-i\left(\mathbf{K}_n+\frac{\mathbf{G}_1}{q}\right)\cdot \mathbf{d}+ i\frac{2\pi \delta_{n,q-1}}{3}} d_B(\mathbf{K}'_{n+1})
\end{equation}
\begin{equation}
\mathbb{R}_{\pi} d_B(\mathbf{K}_n) \mathbb{R}_{\pi}^\dagger=e^{-i\left(\mathbf{K}_n\right)\cdot \mathbf{d}} d_A(\mathbf{K}'_{n+1})
\end{equation}
\end{subequations}
Now for the $\frac{2\pi}{3}$ rotation
$\mathbb{R}_{\frac{2\pi}{3}}$. This is the most complicated of all
because it maps a particular Dirac point to a linear combination of
all Dirac points with the same winding number. The results we quote
below are empirical in the sense that we have not been able to prove
them: rather, we fitted the action of $\mathbb{R}_{\frac{2\pi}{3}}$ on
the Rhim-Park wavefunctions to an analytic form, and checked them for
many values of $q$.

\begin{subequations}
\begin{equation}
\mathbb{R}_{\frac{2\pi}{3}} d_A(\mathbf{K}_n) \mathbb{R}_{\frac{2\pi}{3}}^\dagger=\frac{1}{\sqrt{q}}\sum_{n'} e^{i \phi_A^{\mathbb{R}_{\frac{2\pi}{3}}}(n,n')}d_A({\mathbf{K}_{n'}})
\end{equation}
\begin{equation}
\mathbb{R}_{\frac{2\pi}{3}} d_B(\mathbf{K}_n) \mathbb{R}_{\frac{2\pi}{3}}^\dagger=\frac{1}{\sqrt{q}}\sum_{n'} e^{i \phi_B^{\mathbb{R}_{\frac{2\pi}{3}}}(n,n')}d_B({\mathbf{K}_{n'}})
\end{equation}
\end{subequations}
where,
\begin{subequations}
\begin{align}
\phi_A^{\mathbb{R}_{\frac{2\pi}{3}}}(n,n')=&\frac{\pi}{12 q}\bigg[(4-5q+q^2)+24n n' \nonumber\\
&+6(n^2+n'^2)-6(q-2)(n+n')\bigg]
\end{align}
\begin{align}
\phi_B^{\mathbb{R}_{\frac{2\pi}{3}}}(n,n')=&\frac{\pi}{12 q}\bigg[(4-13q+q^2)+24n n' \nonumber\\
&+6(n^2+n'^2)+(4-6q)n+(20-6q)n'\bigg]
\end{align}
\end{subequations}

Finally the action of the  chiral symmetry on the low energy subspace.
\begin{subequations}
\begin{align}
\mathbb{S} d_A(\mathbf{K}_n) \mathbb{S}^{-1}=d_A^\dagger(\mathbf{K}_n)
\end{align}
\begin{align}
\mathbb{S} d_B(\mathbf{K}_n) \mathbb{S}^{-1}=-d_B^\dagger(\mathbf{K}_n)
\end{align}
\end{subequations}

\bibliography{graphene}

\end{document}